\documentclass[]{aastex63}

\usepackage{changepage}
\usepackage{appendix}
\usepackage[graphicx]{realboxes}
\usepackage{subfigure}
\usepackage{amsmath}
\usepackage{natbib}

\newcommand{\IPAC}{\affil{Caltech-IPAC, 1200 E. California Blvd. Pasadena, CA 91125, USA}}

\newcommand{\STScI}{\affil{Space Telescope Science Institute, 3700 San Martin Drive, Baltimore, MD 21218, USA}}

\newcommand{\STEWARD}{\affil{Steward Observatory, Department of Astronomy, University of Arizona, 933 N. Cherry Ave, Tucson, AZ 85721, USA}}
\received{\today}
\revised{\today}
\accepted{\today}

\usepackage{threeparttable}

\submitjournal{ApJ}

\shorttitle{HeII Dust}
\shortauthors{Maschmann et al.}

\begin{document}

\title{Testing He\,II Emission from Wolf-Rayet Stars as a Dust Attenuation Measure in eight Nearby Star-forming Galaxies}

\correspondingauthor{Daniel Maschmann}
\email{danielmaschmann@arizona.edu}

\author[0000-0001-6038-9511]{Daniel Maschmann}
\STEWARD

\author[0000-0003-2685-4488]{Claus Leitherer}
\STScI

\author[0000-0002-9382-9832]{Andreas L. Faisst}
\IPAC

\author[0000-0002-2278-9407]{Janice C. Lee}
\STScI
\STEWARD

\author{Rebecca Minsley}
\STEWARD

% \correspondingauthor}
% \email{}

\begin{abstract}
% background
The ability to robustly determine galaxy properties such as masses, ages and star-formation rates is critically limited by the ability to accurately measure dust attenuation. 
Dust reddening is often characterized by comparing observations to models of either nebular recombination-lines or the ultra violet (UV) continuum. 
% what we are doing
Here, we use a new technique to measure dust reddening by exploiting the He\,II\,$\lambda$1640 to $\lambda$4686 emission lines originating from the stellar winds of Wolf-Rayet stars. 
The intrinsic line ratio is determined by atomic physics, enabling an estimate of the stellar reddening similar to how the Balmer lines probe reddening of gas emission. 
The He\,II line ratio is measured from UV and optical spectroscopy using the Space Telescope Imaging Spectrograph (STIS) on board the Hubble Space Telescope (HST) for eight nearby galaxies hosting young massive star clusters.
% results
We compare our results to dust reddening values estimated from UV spectral slopes and from Balmer line ratios and find tentative evidence for systematic differences. The reddening derived from the He II lines tends to be higher, whereas that from the UV continuum tends to be lower. A larger sample size is needed to confirm this trend. If confirmed, this may indicate an age sequence probing different stages of dust clearing.
Broad He\,II lines have also been detected in galaxies more distant than our sample, providing the opportunity to estimate the dust reddening of the youngest stellar populations out to distances of $\sim$100~Mpc.
\end{abstract}


\section{Introduction}\label{sec:intro}
% introductory words about dust
Dust is a constant companion of star formation, since it is not only a component of the clouds that collapse to stars, but is also re-injected into the interstellar medium (ISM) by feedback mechanisms providing material for the next stars to form. It is therefore of great importance to quantify dust in star forming regions and young star clusters in order to be able to determine properties such as star formation rates, masses, ages or metallicities \citep{calzetti_dust_2009}.
% say the direct aim of this work
Light passing through dust clouds is attenuated, where the intensity of the attenuation depends on the wavelength. Specifically, the bluer part of a spectrum is stronger attenuated compared to the red, hence the spectral energy distributions are reddened. Reddening is quantitatively described by dust attenuation curves \citep[][]{fitzpatrick_average_1986, calzetti_dust_1994, salim_dust_2018}. 
In practice, dust attenuation is measured by comparing observations with model predictions either through the stellar UV continuum \citep[e.g.][]{calzetti_dust_1994} or recombination-line ratios of hydrogen \citep[e.g.][]{dominguez_dust_2013}. 
In this work we aim to test an alternative method to estimate dust reddening in nearby galaxies relying on \textit{stellar} He\,II emission lines.
Two lines are of special interest: He\,II\,$\lambda$1640 and $\lambda$4686 as they can be observed in the local Universe with UV and optical spectroscopy, respectively. These lines are prevalent as nebular recombination lines in the presence of a hard ionizing source (ionization energy of He+ is $54.4$ eV) such as an active galactic nucleus (AGN) or stellar sources such as X-ray binaries in H\,II regions \citep{shirazi_strongly_2012}. They also occur in stellar winds surrounding Wolf-Rayet (W-R) stars. Their large line width due to the high wind velocities, uniquely distinguishes them from nebular emissions \citep{schaerer_about_1996}. 
The stellar He\,II\,$\lambda$1640 and $\lambda$4686 lines can be considered as almost pure recombination lines, and thus their intrinsic flux ratio is determined by atomic processes. This ratio decreases due to dust attenuation enabling a {\it stellar} dust reddening estimate, first proposed by \citet{conti_new_1990}. The theoretical value is predicted to be $8.36$ for an electron density of $n\rm _e = 10^9$ cm$^{-3}$, an electron temperature of $T\rm_e=20\,000$ K and a case B recombination \citep{hummer_recombination-line_1987}, which is representative for stellar winds in W-R stars. \citet{crowther_reduced_2006} estimated the line ratio to be of the order of $10$ from stellar atmosphere models for all types of W-R stars of the nitrogen sequence (WN).
Based on stellar models, \citet{schaerer_new_1998} found an average line ratio of $7.55$ and $7.95$ for late WN stars in the Galaxy and the Large Magellanic Cloud (LMC), respectively.
By using reddening corrected line fluxes, \citet{leitherer_he_2019} empirically estimated an intrinsic He\,II\,$\lambda$1640 to $\lambda$4686 line ratio of $7.76$ from W-R stars in the Galaxy and the Large Magellanic Cloud (LMC). As the latter estimates does not suffer from systematics due to model choices, we adopt this value throughout this work.

%% why Wr stars are a good choice.
%One of the main motivations to use recombination lines emitted by W-R stars as a dust tracer is that this method only probes the youngest stellar populations in a very specific evolutionary phase.
%This is due to the fact that W-R stars only appear in the first ${\rm 5\,Myr}$, whereas the W-R phase of a star lasts for $< 1 {\rm Myr}$ \citep{meynet_stellar_2005}.
%In our Galaxy and the Magellanic Clouds W-R stars are studied extensively by means of their large variety of spectroscopic features \citep{crowther_physical_2007}. 
%The occurrence rate of W-R stars depends on the ISM metallicity \citep{conti_evolution_1983, van_der_hucht_viith_2001}, which also has a great impact on their spectroscopic features \citep{mokiem_empirical_2007}.
%However, it is important to emphasize that the origin of the W-R stars is not important for the present study. They can be the result of a single O-type star evolving into a W-R star \citep{meynet_massive_2017} or through mass transfer in a binary star system \citep{eldridge_binary_2017}. The He\,II ratios are not affected by the type or evolution of the W-R stars, the initial mass function, or the metallicity \citep{leitherer_he_2019}. 
%However, an important requirement of this method is the need of sufficiently high spectral resolution in order to distinguish broad from narrow He\,II lines. 

% why Wr stars are a good choice.
One of the main motivations to use recombination lines emitted by W-R stars as a dust tracer is that this method only probes the youngest stellar populations in a very specific evolutionary phase.
W-R stars mainly appear in the first ${\rm 5\,Myr}$, whereas the W-R phase of a star lasts for $< 1 {\rm Myr}$ \citep{meynet_stellar_2005}.
They can be the result of a single O-type star evolving into a W-R star \citep{meynet_massive_2017} or through mass transfer in a binary star system \citep{eldridge_binary_2017}. 
In metal rich environments (approximately solar metallicity), W-R stars can be produced at later times $\sim {\rm 10\,Myr}$, whereas in metal poor environments (sub-solar metallicity) no W-R stars are expected after ${\rm 5\,Myr}$ \citep[see Figure\,14 in ][]{leitherer_effects_2014}. In this work we can assume the latter case as the galaxy sample has sub-solar metallicity \citep[see Table\,1 in][]{chandar_ngc_2004}. The only exception is the galaxy He\,2-10. However, the estimated age of the region studied here is 5\,Myr \citep[see Table\,2 in][]{chandar_ngc_2004}.
In our Galaxy and the Magellanic Clouds W-R stars are studied extensively by means of their large variety of spectroscopic features \citep{crowther_physical_2007}. 
The occurrence rate of W-R stars depends on the ISM metallicity \citep{conti_evolution_1983, van_der_hucht_viith_2001}, which also has a great impact on their spectroscopic features \citep{mokiem_empirical_2007}.
It is important to emphasize that the origin of the W-R stars is not important for the present study. The He\,II ratios are not affected by the type or evolution of the W-R stars, the initial mass function, or the metallicity \citep{leitherer_he_2019}. 
However, an important requirement of this method is the need of sufficiently high spectral resolution in order to distinguish broad from narrow He\,II lines. 

% WR time scales and helium dwarfs 
The ages quoted above are based on single-star models. For binary star systems on the other hand, these time scales are longer: 
mass transfer in binaries can lead to hydrogen-free (or -deficient) stars after 5\,Myr. These stars would be classified as W-R \citep{xiao_emission-line_2018}. 
Furthermore, recent discoveries of intermediate mass helium dwarfs \citep{drout_discovery_2023,gotberg_stellar_2023} suggest that there is in fact a continuum of stripped stars from W-R stars to intermediate mass helium dwarfs. However, these stars do not indicate strong emission lines which makes it unlikely that broad He\,II lines trace populations older than $\sim5\,$Myr. Furthermore, the ages of the targets we are using in this study are estimated to be $\leq {\rm 7\,Myr}$ \citep[][their Table\,2]{chandar_ngc_2004}.

% strategy of our method here
We have chosen a sample of eight galaxies hosting bright and massive young star clusters. 
We observed these sources with the Space Telescope Imaging Spectrograph (STIS), on board the Hubble Space Telescope (HST) at UV and optical wavelengths using narrow apertures ($0.2^{\prime\prime}$ and $0.5^{\prime\prime}$). 
These observations also enable us to use the UV-slope as a stellar dust estimator and compare our results. In addition, we use Balmer line measurements from the literature to compare dust-reddening values of stellar and nebula origin.

% structure of this paper
The remainder of this paper is organized as follows. In Section~\ref{sect:heii_obs} we describe the nearby galaxy sample targeted for UV-optical spectroscopy with HST-STIS.  We also summarize the observations, and describe the steps taken to reduce the data.  In Section~\ref{sect:balmer_line}, we describe existing data used in our analysis: spectroscopy from VLT Multi Unit Spectroscopic Explorer (MUSE) and the Sloan Digital Sky Survey (SDSS), as well as measurements reported in the literature.  In Section~\ref{sect:reddening}, the procedures used to calculate E(B-V) from the He\,II line ratio, Balmer line ratio, and UV continuum are discussed.  In Section~\ref{sect:discussion}, we compare the reddening as computed from all three methods, and find general agreement. We discuss the potential for measuring dust attenuation using the He\,II line ratio in higher redshift galaxies.  Finally, we provide a summary of our analysis and results in Section~\ref{sect:conclusion}.

\section{Observations}\label{sect:heii_obs}
% short summary on what awaits the reader in this section
To measure dust attenuation through broad He\,II lines outside the Local Group of galaxies, high-quality spectroscopic observations in the UV and optical wavelength range are needed. 
In addition, both observations must have co-spatial apertures as to probe the emission at the same spatial location.
Therefore, the target sample has to be selected from known extragalactic nearby H\,II regions containing W-R stars, and the spectra used for the subsequent analysis need to be co-spatial.

\subsection{Target Sample}\label{ssect:sample}
% the main idea of the sample
For this study, we selected star-forming regions in eight nearby galaxies (${ 0.0010 \lesssim z \lesssim 0.0095}$, at distances of 3-37\,Mpc) with existing He\,II\,$\lambda$1640 HST-STIS UV observations from the Mikulski Archive for Space Telescopes (MAST). These UV spectra were previously obtained for He 2-10, NGC 3125, Mrk 33, NGC 4214, NGC 4670, and Tol 1924-416 as part of program 9036 (PI: C. Leitherer) taken in 2001–2003, and for NGC 3049 and Tol 89 as part of program 7513 (PI: C. Leitherer) observed in 1999-2000. 
Additional optical HST-STIS observations, providing He\,II\,$\lambda$4686 line measurements were obtained from the HST program GO-15846 (PI: C. Leitherer), executed in 2020-21. These optical observations were carried out at the exact same location, same orientation, and with the same aperture sizes as the UV observations. This is crucial for comparing the UV and optical measurements of the same stellar regions in the galaxies. Furthermore, space-based observations are invaluable for reaching the necessary spatial resolution. A detailed description of the reduction of these new data will be provided in Section~\ref{sec:stis_data_reduction}. 

% What have been done so far with the spectra
The analysis of galaxies in program 9036 was published in \citet{chandar_ngc_2004} and \citet{chandar_stellar_2005}, quantifying the W-R and O-star content for selected massive young star clusters and to better understand strong He~II~$\lambda$1640 features observed in Lyman break galaxies (LBGs) at redshifts $z\sim3$.
The analysis of the UV spectra for NGC 3049 was published in \cite{gonzalez_delgado_massive_2002} with the goal of studying massive stars in metal-rich starbursts and performing consistency tests of existing starburst models for such high-metallicity environment.
The giant H\,II region Tol 89 in NGC 5398 was analyzed in \cite{sidoli_massive_2006} with supplementary HST imaging and Very Large Telescope (VLT)/UV–Visual Echelle Spectrograph (UVES) spectroscopy. They resolve the substructure of this region and identified super star clusters based on spectral modelling.  
These eight selected galaxies are widely studied and well known for their young massive star-forming regions and therefore form an excellent sample to study their He\,II emission line properties.
In Table~\ref{table:targets}, we present the most important characteristics such as coordinates, redshift, and Galactic extinction values, and the STIS aperture widths for the galaxies.
\begin{table}[htp]
%\centering
%\begin{tiny}
\begin{center}
\caption{Galaxy Sample \label{table:targets}}
\begin{tabular}{lccccccc}
\hline\hline
\multicolumn{1}{c}{Galaxy} & 
\multicolumn{1}{c}{R.A.} &
\multicolumn{1}{c}{DEC.} &
\multicolumn{1}{c}{$z$} & 
\multicolumn{1}{c}{$A_{\rm V}$ (gal)} &
\multicolumn{1}{c}{Slit Width }&
\multicolumn{2}{c}{Total exp} \\
& & & & & & \multicolumn{1}{c}{G140L} & \multicolumn{1}{c}{G430M} \\
& & & & \multicolumn{1}{c}{mag} & \multicolumn{1}{c}{$\arcsec$} & \multicolumn{1}{c}{s} & \multicolumn{1}{c}{s} \\
\hline
He 2-10      & 08h36m15.13s & $-$26$^{\circ}$24m33.7s & 0.0029 & 0.31 & 0.2 & 10891 & 4195 \\
NGC 3049     & 09h54m49.40s & +09$^{\circ}$16m15.9s & 0.0049 & 0.11 & 0.5 & 11064 & 1650 \\
NGC 3125     & 10h06m33.29s & $-$29$^{\circ}$56m06.8s & 0.0037 & 0.21 & 0.2 & 3975 & 4084 \\
Mrk 33       & 10h32m31.88s & +54$^{\circ}$24m02.2s & 0.0048 & 0.03 & 0.2 & 1446 & 4084 \\
NGC 4214     & 12h15m39.45s & $+$36$^{\circ}$19m34.8s & 0.0010 & 0.06 & 0.2 & 1544 & 4235 \\
NGC 4670     & 12h45m17.44s & +27$^{\circ}$07m31.8s & 0.0036 & 0.04 & 0.2 & 1470 & 4090 \\
Tol 89       & 14h01m19.92s & $-$33$^{\circ}$04m10.7s & 0.0041 & 0.18 & 0.5 & 11620 & 1650 \\
Tol 1924-416 & 19h27m58.31s & $-$41$^{\circ}$34m29.8s & 0.0095 & 0.24 & 0.2 & 4044 & 4108 \\  
\hline
\end{tabular} 
\end{center}
%\end{tiny}
\tablecomments{Redshifts and Galactic $A\rm_V$ are taken from the NASA$/$ IPAC Extragalactic Database. We further list the STIS slit widths and the exposure times.}
\end{table}
%

% what the data will provide and a justification of the selected sources
The present work is a pilot program to test the application of He~II lines as dust tracers in nearby galaxies, thus, the sources were selected as the strongest He~II emitters. In particular, NGC~3125 is known to be the strongest He~II~$\lambda$1640 source in the local Universe \citep{chandar_ngc_2004} and therefore provides excellent conditions to test these mechanisms.
Furthermore, some of the selected targets have multiple sources covered with STIS slit pointing. He~2-10 contains four strong sources, with a sufficient spatial separation for an individual spectral analysis of each of them. Mrk~33, NGC~3049 and NGC~4670 also have multiple sources, but as described in Section~\ref{sec:stis_data_reduction}, the measured He\,II suffer from low signal-to-noise (S/N) values and only one source will ultimately provide a He\,II\,$\lambda$1640, $\lambda$4686 line ratio.

\subsection{HST-STIS Observations}\label{sect:stis_obs}
% describing the observation on a technical level.
The UV HST-STIS spectra were taken in the far-UV using the Multianode Microchannel Array (FUV-MAMA) detector with the STIS G140L grating, optical spectra are taken with the G430M gratings with the STIS/CCD. 
The G140L grating covers a wavelength range of 1150 to 1730 ${\rm \AA}$, with a dispersion of $0.6\,{\rm \AA \, pixel^{-1}}$ and a pixel scale of $\sim0.025\,^{\prime\prime}\,{\rm pixel^{-1}}$.
Spectra obtained with the grating G430M cover a wavelength range and dispersion of 4563 to 4849 ${\rm \AA}$ and $0.28\,{\rm \AA \, pixel^{-1}}$, respectively. The pixel scale for all optical spectra is $\sim~0.051\,{\rm \arcsec/pixel}$.

\begin{figure*}[h]
    \centering
    \includegraphics[width=0.95\textwidth]{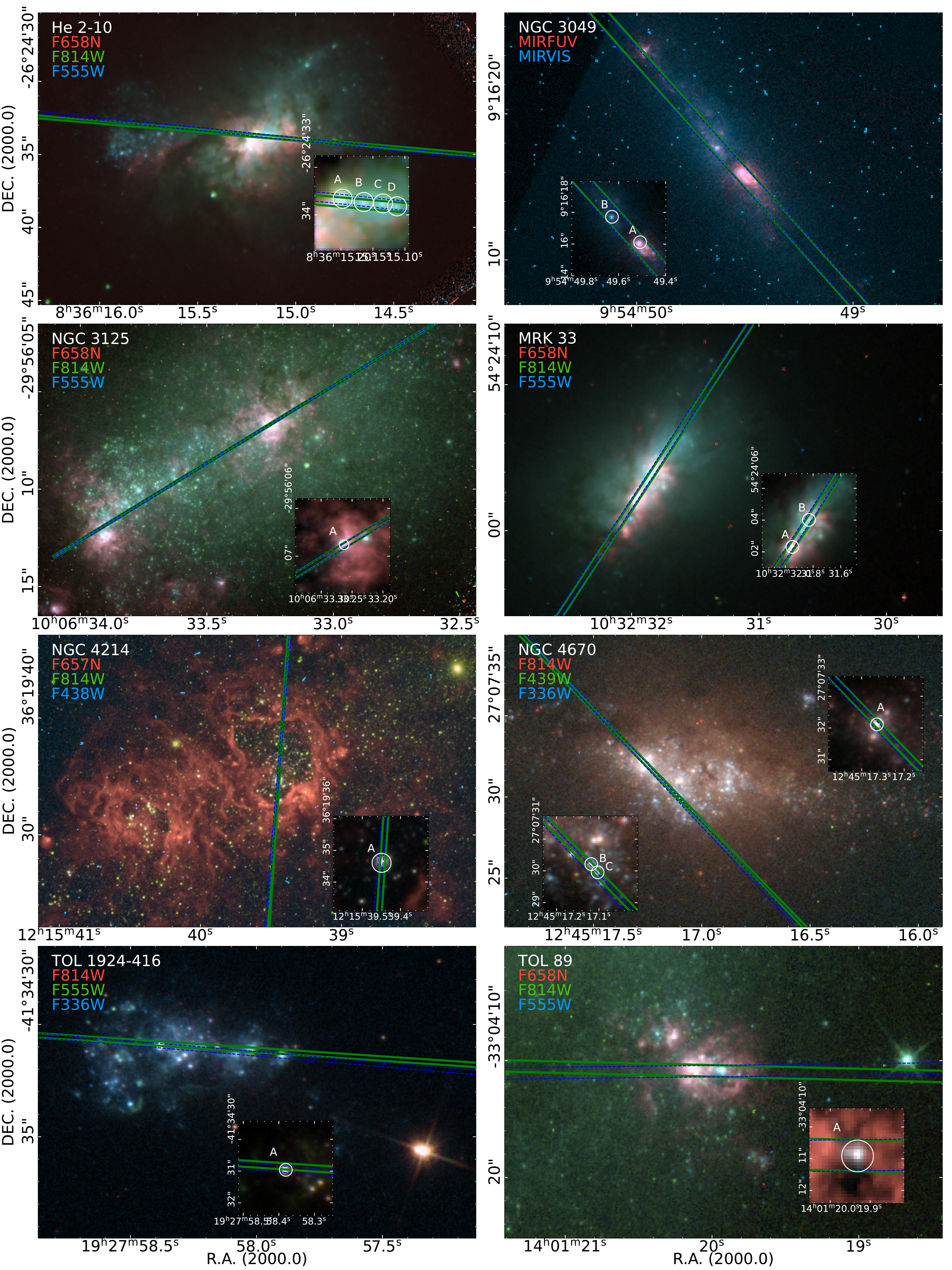}
    %\caption{Observed galaxies with STIS slit positions overlaid. For each galaxy we show HST composite images taken from the archive using the software \textsc{multicolorfits}\footnote{\url{https://github.com/pjcigan/multicolorfits}} and include the aligned STIS slits with a solid green (resp. dashed blue) line for the G140L (resp. G430M) grating. The zoom-in panels show the identified He\,II emitters.} 
    \caption{Observed galaxies with STIS slit positions overlaid. For each galaxy we show HST composite images taken from the archive using the software  \textsc{multicolorfits}$^{\rm a}$ and include the aligned STIS slits with a solid green (resp. dashed blue) line for the G140L (resp. G430M) grating. The zoom-in panels show the identified He~II emitters.} 
    \label{fig:slit_alignment}
    \begin{tablenotes}
    \item $^{\rm a}$ \url{https://github.com/pjcigan/multicolorfits}
    \end{tablenotes}
\end{figure*}

A perquisite for our analysis is that our observations are co-spatial at UV {\it and} optical wavelengths. At the outset of our analysis we therefore confirm that the orientation of the slit is identical across the three gratings for all targets. The STIS instrument target acquisition process is done in three main parts. The first two involve centering the target in a $100 \times 100$ pixel target acquisition sub-array. After the initial guide star acquisition, the instrument makes an initial pointing whereby the target is captured within the acquisition sub-array. The instrument then does a coarse centering in which the target is placed with respect to a reference point in the target acquisition sub-array. The third step performs the fine centering, which places the object precisely within the slit.
In Figure~\ref{fig:slit_alignment}, we show the slit positions of both observations superposed on existing HST image observations.

\subsection{STIS Data Reduction}\label{sec:stis_data_reduction}
% say what is the plan in this section 
We identify the spectral traces of massive star-forming regions on the observed 2-D STIS images and convert them to 1-D spectra. The goal is to measure the He\,II\,$\lambda$1640 and $\lambda$4686 line fluxes in the G140L and G430M spectra, respectively, on which a subsequent dust attenuation estimate is based. Furthermore, we estimate the slope of the UV continuum emission in the G140L spectra, providing an independent measurement of the stellar dust attenuation. 

% 2D data reduction
All STIS data are retrieved from the MAST and further processed using the {\it CALSTIS} pipeline \citep[see][]{sohn_stis_2019}. 
The basic two-dimensional image processing steps done by the pipeline differ slightly depending on the detector used to observe the data (MAMA or CCD). The basic 2-D image module (\textit{basic2d}) consists of the following tasks: bad pixel flagging, conversion of native high resolution pixels to low resolution pixels (MAMA), global linearity correction (MAMA), overscan and bias subtraction (CCD), dark subtraction, flat-field corrections and wavelength calibration. 
The pipeline also includes cosmic-ray rejection used for CCD recorded data. 
For targets with multiple exposures, we check for any spatial offsets between exposures (none were found to have $>0.25\,{\rm pixel}$ offsets) and co-add the raw data files before processing them with the {\it CALSTIS} pipeline.
The total exposure times across all observations are given in Table\,\ref{table:targets}. 

\begin{figure*}[h!]
    \centering
    \includegraphics[width=0.95\textwidth]{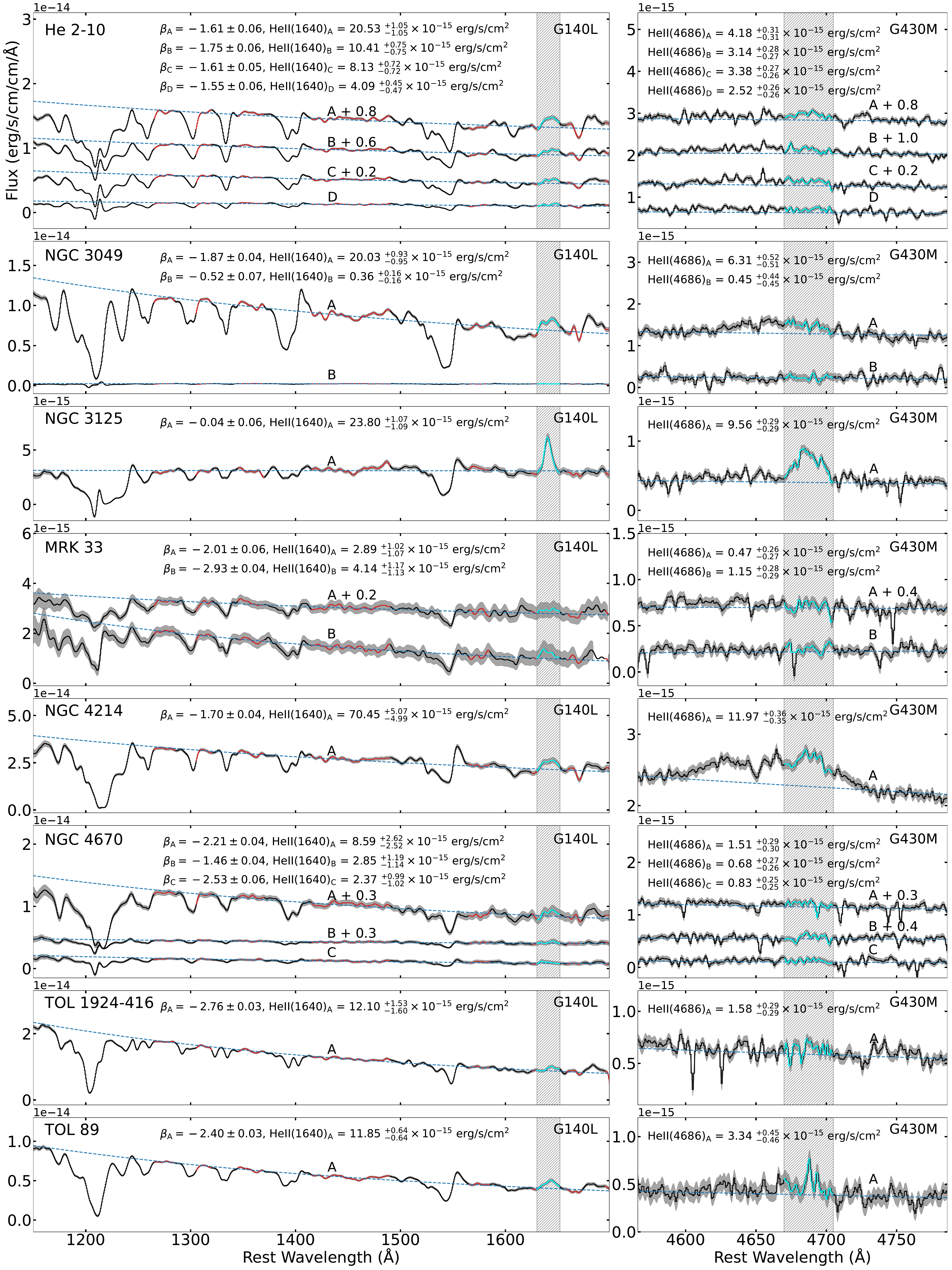}
    \caption{Extracted STIS spectra for the G140L grating and the He\,II\,$\lambda$4686 line in the G430M grating. For each galaxy, we display the spectra of all identified He~II emitters. For multiple spectra we add an offset to distinguish them. We show the continuum fit with blue dashed lines. The data points used for the continuum fit for the G140L grating are marked in red. The wavelength windows for the He\,II\,$\lambda$1640 and $\lambda$4686 lines are marked with a gray background and the lines are highlighted in turquoise. The measured quantities for each spectrum are displayed on the top and also summarized in Table\,\ref{table:ebv}.} 
    \label{fig:stis_spec}
\end{figure*}

% getting the 1d spectra
The 2-D spectral images are used to identify the spectral traces of the bright star-forming regions. The width of the extraction box is identified by eye to include the total flux and at the same time minimize contamination from neighboring sources. The extraction box pixel sizes are slightly larger for the UV spectra (9-19 pixels), in comparison to the optical spectra (5-9 pixels) due to the smaller pixel size of the STIS FUV-MAMA detector compared to that of the CCD detector. 
We use the {\it x1d} ststools function to extract the one-dimensional data. This function automatically calculates and subtracts the background, corrects the wavelengths to a heliocentric reference frame, and does the absolute flux calibration. We then convert the observed wavelength to rest-frame wavelength using the redshift measurement from the spectrum. 
The G140L spectra and the He\,II\,$\lambda$1640, $\lambda$4686 lines are shown for each identified source in Figure~\ref{fig:stis_spec}.
As a final step, all spectra are visually inspected to identify and manually remove detector flaws. This is for example the case for the G430M spectrum of the target Mrk~33-B at around 4675~\AA\ rest-frame wavelength. A cosmic ray in the flat field was not detected by the pipeline and therefore was not properly flagged. In this case, the corresponding wavelength was flagged and removed in the subsequent analysis.
%

% measurement of the UV slope 
We first measure the UV continuum slope $\beta$ by fitting a linear function to the data points between 1240 and 1700~${\rm \AA}$. In order to avoid absorption and emission features, the fit only considers specific wavelength intervals as specified in \citet{calzetti_dust_1994}. 
We de-redden the flux to account for Milky Way extinction using the \cite{cardelli_relationship_1989} reddening law before measuring the UV slope.
The best fit to the continuum data points are shown for each spectrum in Figure~\ref{fig:stis_spec}, and the measured $\beta$-values are listed in Table\,\ref{table:ebv}. 

% measuring the Heii flux
Subsequently, we use the estimated UV continuum to determine the He\,II\,$\lambda$1640 line flux from the continuum subtracted spectra. To measure the He\,II\,$\lambda$4686 line flux, we subtract the continuum, estimated with a linear fit between 4600 and 4800 ${\rm \AA}$, excluding regions with known emission lines. 
The line fluxes are measured by summing up the flux between 1631 and 1652 ${\rm \AA}$ for the He\,II\,$\lambda$1640 line and between 4670 and 4705 ${\rm \AA}$ for the He\,II\,$\lambda$4686 line. Note that the latter blended with [Fe\,III]\,$\lambda4658$ and other stellar lines, whose ratios depend on the types of W-R stars present. The wavelength range is chosen based on the width of the He\,II\,$\lambda$1640, which is expected to be comparable to the width of He\,II\,$\lambda$4686 \citep{leitherer_he_2019,hillier_empirical_1987}.
We estimate the line uncertainties with a Monte-Carlo approach by randomly adding Gaussian noise with the scale of the measured noise level in the selected region. We repeat this 1000 times and compute the uncertainties as the 68-percentile.

\section{Balmer Line Measurements}\label{sect:balmer_line}
In addition to reddening estimates based on He\,II lines and the UV slope, alternative tracers can be used for comparison. Balmer recombination lines emitted from H~II regions are one of the most commonly used tools for reddening estimates \citep[e.g.][]{dominguez_dust_2013,momcheva_nebular_2013}. 
For the present sample, we collected Balmer line observations from the VLT-MUSE, SDSS and individual spectroscopic estimates from the literature.

\subsection{Archival MUSE Observations}\label{sssect:muse}
Due to its high spatial resolution, MUSE observations are well suited to estimate Balmer line fluxes for the regions observed with STIS.
We selected archival MUSE observation performed in the wide field mode from the ESO-Archives for He\,2-10 (id: 095.B-0321), NGC\,3125 (id: 094.B-0745) and Tol\,1924-416 (id: 60.A-9314). 
The spatial resolution for each target is below one arc-second (taking the individual seeing into account) and covers the optical wavelengths from 4650 to 9300~\AA, which includes the H$\beta$ and H$\alpha$ lines.
To identify the correct region of He\,II emitters, we aligned the MUSE observations with the HST imaging observations, which we had previously aligned with the STIS observations (See Section~\ref{sect:heii_obs} and Figure\,\ref{fig:slit_alignment}). In order to properly align the MUSE data cube, we cross-identified three to six point-like bright clusters or H~II regions in both, MUSE and HST imaging, and calculated a new solution for the world coordinate system of the MUSE observation. This alignment is crucial to select the same stellar regions as with the HST-STIS observations.

Considering the STIS slit width of $0.2\arcsec$, the spatial resolution of MUSE is insufficient to properly resolve multiple sources which are aligned along the STIS slit as we find for He\,2-10. In this case all four identified He\,II emitters are located within $1\arcsec$. Even though the spectra are cross-contaminated due to the PSF and the seeing, we select each source individually which is made possible by the MUSE pixel size of $0.2\arcsec\times0.2\arcsec$.

The He\,II emitter in the galaxy Tol\,1924-416 is fairly isolated and has a diameter of $0.5\arcsec$, excluding most possible sources of contamination.
For the H~II region in NGC\,3125 on the other hand the spectral selection is not so straight forward. In fact this region is so bright, that in the central part of the region the H$\alpha$ emission is saturated in the MUSE observations. In order to estimate the Balmer lines from this regions, we only select the surrounding non-saturated pixels within a radius of $0.5^{\prime\prime}$.

\subsection{Archival SDSS Spectra}\label{sssect:sdss}
Spectroscopic observations from SDSS are integrated over a circular aperture of $3^{\prime\prime}$ and cover the wavelength range from 3800 to $9200\,{\rm \AA}$, enabling the measurements of the H$\gamma$, H$\beta$ and H$\alpha$ lines. 
We find SDSS spectral observations for Mrk\,33, NGC\,3049 and NGC\,4670 centered on the massive star formation regions, which are among our targets. 
Despite the difference in aperture between the STIS and the SDSS observations, these provide a good estimate assuming most of the nebular emission comes from the selected young star cluster regions. For the galaxy Mrk\,33, the H$\alpha$ emission line is saturated in the SDSS spectra, and we therefore use the H$\beta$ and H$\gamma$ ratio to estimate the Balmer decrement.

\subsection{Balmer line observations in the literature}\label{sssect:literature}
The massive star-forming region Tol\,89 in NGC\,5398 was observed with the VLT-UVES by \citet{sidoli_massive_2006}. These observations are co-spatial with the STIS observations and therefore provide a precise measurement for the Balmer decrement estimated from their estimated H$\alpha$ and H$\beta$ line fluxes.

For NGC\,4214, we use archival spectra provided by \citet{moustakas_integrated_2006}, who observed this target with a long slit by drifting over the entire galaxy. These observations only provide a global estimate of the Balmer decrement which can be seen as an average of the galaxy and not as representative of the relatively small region probed by the STIS observations.

\section{Reddening estimates}\label{sect:reddening}
Dust in the interstellar medium causes wavelength-dependent attenuation. Specifically, the slope of the stellar UV continuum and the ratio of emission lines situated at different wavelengths change. Recombination lines, such as the Balmer line series, have a known intrinsic ratio determined by atomic physics and $n{\rm_e}$ and $T{\rm_e}$ of the H~II region.  
The novelty of this work is to test the estimate of dust attenuation based on \textit{stellar} (as opposed to nebular) He\,II lines, calibrated in  \cite{leitherer_he_2019}. In order to evaluate this method we compare the resulting {\it E(B-V)} values to estimates based on the stellar UV-continuum and nebular Balmer lines.

\subsection{Reddening estimates based on He\,II\,$\lambda$1640 and $\lambda$4686}\label{sect:red_heii}
The ratio between the stellar He\,II\,$\lambda$1640 and $\lambda$4686 lines is a fixed value. Following \citet{leitherer_he_2019} the intrinsic ratio is 
\begin{equation}
    R_0 = \frac{F_0(1640)}{F_0(4686)},
\end{equation}
where $F_0(1640)$ and $F_0(4686)$ are the reddening-free line fluxes.
Observations of this ratio ($R_{\mathrm{obs}}$) decrease from the effects of dust attenuation. 
The observed, reddened and the intrinsic, unreddened ratio are related as, 
\begin{equation}
    \frac{R_{\mathrm{obs}}}{R_{\mathrm{0}}} = 10^{0.4\, E(B-V) [k(4686) - k(1640)]}\, , \label{eq:2}
\end{equation}
where $k(1640)$ and $k(4686)$ are the absorption coefficients at 1640 and 4686 ${\rm \AA}$ respectively. Rearranging Equation~\ref{eq:2} for $E(B-V)$ gives
\begin{equation}\label{eq:3}
    E(B-V) = \frac{1}{0.4 [k(4686) - k(1640)]}\log R_{\mathrm{obs}} + \frac{1}{0.4 [k(1640) - k(4686)]}\log R_{0}.  
\end{equation}
Here, we use the intrinsic ratio of $\log R_0 = 0.89$ derived by \citet{leitherer_he_2019}. We adopt the parameterization of \citet{reddy_mosdef_2015} who expressed $k(\lambda)$ as a 3rd-order polynomial
\begin{equation}\label{eq:4}
    k(\lambda) = -5.726 + \frac{4.004}{\lambda} - \frac{0.525}{\lambda^2} + \frac{0.029}{\lambda^3} + 2.505,
\end{equation}
providing the absorption coefficients $k(1640)=8.249$ and $k(4686)=3.215$. 
Applied to Equation~\ref{eq:3}, we obtain the relation 
\begin{equation}\label{eq:5}
    E(B-V)_{\mathrm{He\, II}} = - 0.497 \log R_{\mathrm{obs}} + 0.442.
\end{equation}

The measured He\,II ratios and the estimated ${E(B-V)_{\rm He\,II}}$ values are listed in Table\,\ref{table:ebv}.

\subsection{Reddening estimates based on UV-slope}\label{sect:red_uv}
The attenuation values derived from the helium lines suggest that the UV wavelengths are strongly affected by dust. This makes the UV continuum a useful estimator of dust attenuation. The wavelength dependence of the UV continuum  is well described by a power law. By measuring the spectral index $\beta$ and comparing it to the intrinsic value, one can directly measure the dust attenuation ${E(B-V)_{\rm UV}}$. \citet{reddy_mosdef_2015} assumed an intrinsic continuum slope of $-2.44$ and found a relation between the UV dust attenuation and $\beta$ of
\begin{equation}\label{eq:6}
    E(B-V)_{\mathrm{UV\, Slope}} = \frac{\beta + 2.44}{4.54}.
\end{equation}
The measurement of the UV power-law indices for the STIS-observed star-forming regions is described in Section~\ref{sect:stis_obs} and the $\beta$, as well as the  ${E(B-V)_{\rm UV}}$ values, are listed in Table~\ref{table:ebv}.

\subsection{Reddening estimates based on Balmer decrement}\label{sect:red_balmer}
Following the same principle as for the He\,II lines, the Balmer lines can be used to measure the dust attenuation. 
W-R emission lines and the UV continuum measured with $\beta$ probe the attenuation of the stellar light, whereas Balmer lines are of nebular origin, hence they probe dust attenuation in nebular regions. Furthermore, the Balmer line measurements are not always taken in the same aperture as the STIS observations, as described in detail in Section~\ref{sect:balmer_line}, and a comparison should therefore be taken with care.

To estimate the dust attenuation, we preferably use the ratio between the H$\alpha$ and H$\beta$ line. However, for the galaxy Mrk\,33, the H$\alpha$ line is saturated in the SDSS spectra, and we therefore use the line ratio between H$\beta$ and H$\gamma$. 
For the dust attenuation we here assume an electron temperature of $T{\rm _e=10^4\,K}$ and an electron density of $n_{\rm e} = 10^2 {\rm cm^{-3}}$ for a case B recombination as described in \citet{osterbrock_astrophysics_1989}. 
Following \citet{momcheva_nebular_2013} and \citet{dominguez_dust_2013}, the dust extinction estimate is based on the assumption of an intrinsic H$\alpha$/H$\beta$ (or  H$\beta$/H$\gamma$) ratio of $2.86$ (or $2.13$) leading to 
\begin{equation}\label{eq:ebv_alpha_beta}
    E(B-V)_{\alpha,\beta} = {\rm 1.97\, log \left[ \frac{(H\alpha / H\beta)_{obs}}{2.86}\right]},
\end{equation}
and
\begin{equation}\label{eq:ebv_beta_gama}
    E(B-V)_{\beta,\gamma} = {\rm 4.43\, log \left[ \frac{(H\beta / H\gamma)_{obs}}{2.13}\right]}.
\end{equation}
The ${E(B-V)_{\rm Balmer}}$ values are listed in Table~\ref{table:ebv}.
\begin{table*}[htp]
\begin{center}
\caption{E(B-V) Measurements \label{table:ebv}}
\begin{tabular}{lcccccccc}
\hline\hline
\multicolumn{1}{c}{Target} & \multicolumn{1}{c}{} &
\multicolumn{1}{c}{$F(1640)$} & \multicolumn{1}{c}{$F(4686)$} & \multicolumn{1}{c}{$F1640/F4686$} & 
\multicolumn{1}{c}{$\beta$} & 
\multicolumn{1}{c}{$E(B-V)_{\mathrm{He\,II}}$} &
\multicolumn{1}{c}{$E(B-V)_{\mathrm{UV\,Slope}}$} &
\multicolumn{1}{c}{$E(B-V)_{\mathrm{Balmer}}$} 
\\ 
\multicolumn{1}{c}{} & \multicolumn{1}{c}{} &
\multicolumn{2}{c}{[$10^{-15} \mathrm{erg}\, \mathrm{s}^{-1} \mathrm{ cm}^{-2}$]} & 
\multicolumn{1}{c}{} & 
\multicolumn{1}{c}{} & 
\multicolumn{1}{c}{mag} &
\multicolumn{1}{c}{mag} &
\multicolumn{1}{c}{mag}
\\
\hline
HE 2-10 & A & 20.5$\pm$1.0 & 4.2$\pm$0.3 & 4.9$\pm$0.4 & -1.6$\pm$0.1 & 0.10$\pm$0.02 & 0.18$\pm$0.01 & 0.13$\pm$0.01 \\ 
 & B & 10.4$\pm$0.7 & 3.1$\pm$0.3 & 3.3$\pm$0.3 & -1.8$\pm$0.1 & 0.18$\pm$0.02 & 0.15$\pm$0.01 & 0.17$\pm$0.01 \\ 
 & C & 8.1$\pm$0.7 & 3.4$\pm$0.3 & 2.4$\pm$0.2 & -1.6$\pm$0.1 & 0.25$\pm$0.02 & 0.18$\pm$0.01 & 0.12$\pm$0.01 \\ 
 & D & 4.1$\pm$0.5 & 2.5$\pm$0.3 & 1.6$\pm$0.2 & -1.6$\pm$0.1 & 0.34$\pm$0.03 & 0.20$\pm$0.01 & 0.22$\pm$0.01 \\ 
NGC 3049 & A & 20.0$\pm$0.9 & 6.3$\pm$0.5 & 3.2$\pm$0.3 & -1.9$\pm$0.0 & 0.19$\pm$0.02 & 0.13$\pm$0.01 & 0.29$\pm$0.03 \\ 
 & B & 0.4$\pm$0.2 & 0.5$\pm$0.4 & -- & -0.5$\pm$0.1 & -- & 0.42$\pm$0.02 & 0.29$\pm$0.03 \\ 
NGC 3125 & A & 23.8$\pm$1.1 & 9.6$\pm$0.3 & 2.5$\pm$0.1 & -0.0$\pm$0.1 & 0.25$\pm$0.01 & 0.53$\pm$0.01 & 0.03$\pm$0.01 \\ 
MRK 33 & A & 2.9$\pm$1.0 & 0.5$\pm$0.3 & -- & -2.0$\pm$0.1 & -- & 0.09$\pm$0.01 & 0.01$\pm$0.01 \\ 
 & B & 4.1$\pm$1.1 & 1.2$\pm$0.3 & 3.6$\pm$0.9 & -2.9$\pm$0.0 & 0.17$\pm$0.06 & -0.11$\pm$0.01 & 0.01$\pm$0.01 \\ 
NGC 4214 & A & 70.4$\pm$5.0 & 12.0$\pm$0.4 & 5.9$\pm$0.2 & -1.7$\pm$0.0 & 0.06$\pm$0.01 & 0.16$\pm$0.01 & 0.06$\pm$0.02 \\ 
NGC 4670 & A & 8.6$\pm$2.6 & 1.5$\pm$0.3 & 5.7$\pm$1.2 & -2.2$\pm$0.0 & 0.07$\pm$0.04 & 0.05$\pm$0.01 & 0.15$\pm$0.03 \\ 
 & B & 2.9$\pm$1.2 & 0.7$\pm$0.3 & -- & -1.5$\pm$0.0 & -- & 0.21$\pm$0.01 & 0.15$\pm$0.03 \\ 
 & C & 2.4$\pm$1.0 & 0.8$\pm$0.3 & -- & -2.5$\pm$0.1 & -- & -0.02$\pm$0.01 & 0.15$\pm$0.03 \\ 
TOL 89 & A & 11.9$\pm$0.6 & 3.3$\pm$0.5 & 3.6$\pm$0.5 & -2.4$\pm$0.0 & 0.17$\pm$0.03 & 0.01$\pm$0.01 & 0.07$\pm$0.01 \\ 
TOL 1924-416 & A & 12.1$\pm$1.6 & 1.6$\pm$0.3 & 7.6$\pm$1.4 & -2.8$\pm$0.0 & 0.00$\pm$0.04 & -0.07$\pm$0.01 & -0.069$\pm$0.002 \\ 
 & B & -1.0$\pm$0.5 & 0.7$\pm$0.3 & -- & -1.1$\pm$0.1 & -- & 0.28$\pm$0.02 & -0.069$\pm$0.002 \\ 
\hline
\end{tabular} 
\end{center}
\tablecomments{We show the He\,II\,$\lambda$1640 and $\lambda$4686 line flux, their ratio and the measured UV slope $\beta$ for all STIS spectra. For targets with line fluxes smaller than a signal-to-noise (S/N) $<3$ we do not calculate the line ratio. We furthermore list the calculated ${E(B-V)_{\rm He\,II}}$ and ${E(B-V)_{\rm UV}}$, as well as the dust reddening estimations from Balmer lines ${E(B-V)_{\rm Balmer}}$. As described in Section~\ref{sect:balmer_line}, in some cases no individual ${E(B-V)_{\rm Balmer}}$ value are estimated for multiple targets in one galaxy due to larger spectral apertures. In such a case we use the same Balmer line estimate for all targets within one galaxy.}
\end{table*}

\section{Discussion}\label{sect:discussion}

\subsection{Origin of different results between the three techniques}
\begin{figure*}[h!]
    \centering
    \includegraphics[width=0.95\textwidth]{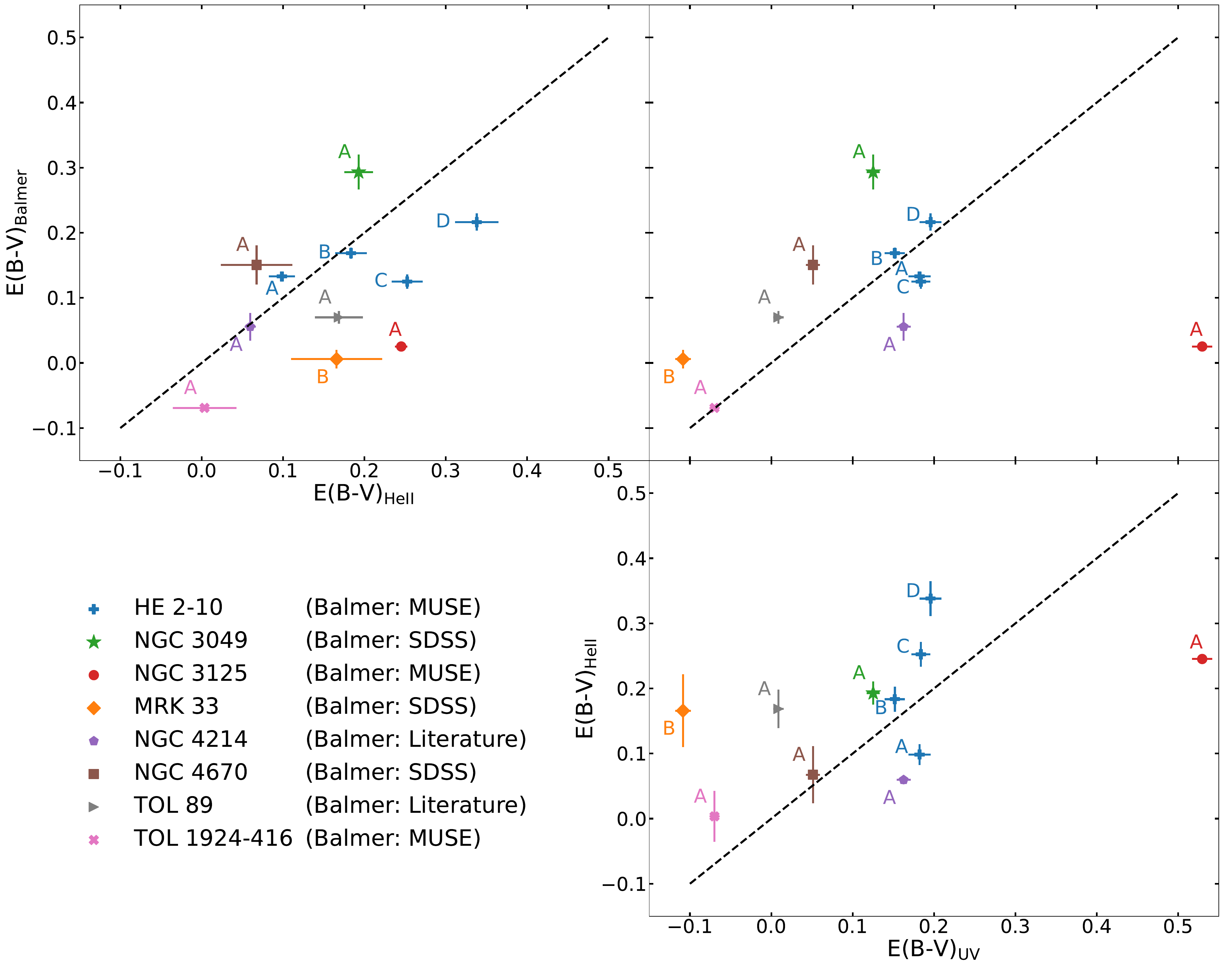}
    \caption{Comparison of all {\it E(B-V)} estimates computed in this work. A dashed line denotes the one-to-one relation. Each galaxy is represented by individual markers and colors. For some galaxies multiple clusters were identified and thus we specify each data point with a letter to cross identify the measurement. The ${E(B-V)_{\rm Balmer}}$ measurements of the target sample are measured with the same method since we collected them from archival observations and literature values. The origin of the Balmer measurements are specified in the legend on the bottom left, and a detailed description of the data estimate is described in Section~\ref{sect:balmer_line}. } 
    \label{fig:compare_ebv}
\end{figure*}

Overall, we find consistency between the different dust reddening estimates used in this work. The comparisons plotted in Figure~\ref{fig:compare_ebv} suggest good agreement between the three methods with the exception of the $E(B-V)_{\rm UV}$ estimate for NGC\,3125. As discussed in the next section, there are reasons for the anomalous $\beta$ slope in this galaxy. Therefore we exclude NGC\,3125 for the calculation of the correlation coefficients of all three combinations and find $E(B-V)_{\rm He\,II} \,vs.\, E(B-V)_{\rm Balmer} = 0.60$, $E(B-V)_{\rm UV} \,vs.\, E(B-V)_{\rm Balmer} = 0.69$ and $E(B-V)_{\rm UV} \,vs.\, E(B-V)_{\rm He\,II} = 0.44$.  Although the statistical power of eight galaxies is limited, the results presented here suggest that the ratio of the He\,II lines can be used as estimators of dust attenuation in star forming regions beyond the Milky Way and the Magellanic Clouds.

However, differences in the determinations are expected due to their different physical origins and the different apertures used to measure ${E(B-V)_{\rm Balmer}}$, ${E(B-V)_{\rm He\,II}}$ and ${E(B-V)_{\rm UV}}$. As discussed in \citet{leitherer_he_2019}, these three determinations of $E(B-V)$ provide a unique opportunity to study the effects of dust at different ages and for the stellar and gas phase separately.  

%Does the dust reddening of the stars change with age from the youngest population, as determined from the
%W-R line ratio, to the older B-star population measured from the UV spectral slope?  Single W-R stars have ages of less than $\sim5\,%{\rm Myr}$ and spend only about $10\%$ of their lifetime in the W-R phase \citep{meynet_stellar_2005}. This gives us a concrete time %constraint as the He\,II lines form in W-R winds and can therefore only be emitted from very young $<5\,{\rm Myr}$ star-formation sites. % In comparison, the UV stellar continuum in young stellar populations is dominated by the photospheric emission from O and B stars. 
%B stars have life times of order $\sim$100\,Myr, which means that the UV continuum probes a different age regime than the He\,II lines %from W-R stars; the far- to near-UV continuum slope measures the dust attenuation of the stellar population at ages of 10 to 100 Myr %\citep{meurer_dust_1999}.
%We may expect ${E(B-V)_{\rm UV}}$ to be lower than ${E(B-V)_{\rm He\,II}}$ as the dust from the natal dust cloud is cleared by winds, %radiation pressure, and supernovae from massive stars.

Is the dust attenuation of the gas, different from that of the stars?  As discussed in \citet{calzetti_dust_1994}, the nebular emission from H~II regions suffers about twice the reddening experienced by the stellar continuum.  This has been confirmed using large samples of nearby galaxies which show that $A$(FUV)/$A$(H$\alpha$) = 1.8 \citep[e.g.,][]{lee_comparison_2009}, as expected for the Calzetti reddening curve and differential extinction law \citep{calzetti_dust_2001}. However, the timescales of these tracers can be different by an order of magnitude, since the nebular Balmer recombination lines probe the dust attenuation of the gas at a characteristic age of $\sim$5 Myr, since only short-lived, massive stars emit sufficient Lyman continuum radiation to create H~II regions.  With ${E(B-V)_{\rm He\,II}}$ we can remove the age as a variable, and ask more directly whether the dust attenuation of the gas, as measured from the Balmer decrement, differs from that of the contemporaneous generation of stars, as measured by the W-R lines.

We calculate the mean reddening derived from the three methods, and obtain:
\begin{itemize}
    \item \;$\overline{E(B-V)}_{\rm He\,II} = 0.15\,\pm\,0.09$  \item \;$\overline{E(B-V)}_{\rm Balmer} = 0.12\,\pm\,0.10$
    \item \;$\overline{E(B-V)}_{\rm UV} = 0.09\,\pm\,0.11$
\end{itemize}
Even though these values agree with each other within their standard deviations, differences are tentatively present, suggesting the sequence
$E(B-V)_{\rm He\,II} > E(B-V)_{\rm Balmer} > E(B-V)_{\rm UV}$. We caution against overinterpreting the significance of this relation, given the size of the statistical errors. In particular, the mean reddening derived from the Balmer lines must be taken with care, as the values were not obtained co-spatially. Ignoring $E(B-V)_{\rm Balmer}$, we can focus on $E(B-V)_{\rm He\,II}$ and $E(B-V)_{\rm UV}$, which are purely {\it stellar} based dust tracers. Then this sequence can be understood by a relatively early onset of the clearing of natal material around the newly formed stars and a short (1-2~Myr) clearing time-scale, consistent with previous studies of star formation timescales \citep[e.g.,][]{whitmore_using_2011,hollyhead_studying_2015, sokal_prevalence_2016,hannon_h_2022}. In fact, the different values would represent a sequence of reddening estimates at different stages of stellar evolution, ranging from $\sim$5~Myr (W-R stars) to $\sim$100~Myr (B stars).

A further aspect which needs to be addressed is the production of dust due to W-R stars. As shown in \citet{lau_revisiting_2020,lau_nested_2022}, carbon-rich W-R (WC) stars are a non-negligible source of dust. However, the He\,II\,$\lambda$1640 and $\lambda$4686 lines are predominantly produced by nitrogen-rich W-R (WN) stars with only a small contribution of WC stars. 
Therefore significant dust production by WC stars appears to be unlikely.

\subsection{Comparison of different reddening in individual galaxies}

In the following, we compare the different reddening estimates for individual galaxies and star-forming regions in our sample. The results are summarized in Figure~\ref{fig:compare_ebv}. We note that for some galaxies not all observed regions have significant He\,II detection. In fact, for all non-detected regions we are unable to compute upper limits for the $E(B-V)_{\rm He\,II}$ values since both He\,II\,$\lambda$1640 and $\lambda$4686 are not detected.

% details about individual galaxies
\paragraph{He\,2-10~}
This is the only galaxy in our sample with multiple detected He\,II emitters aligned in the STIS slit. The individual ${E(B-V)_{\rm He\,II}}$ values range from 0.10\,mag (source A) to 0.34\,mag (source D), suggesting that these 4 clusters are at different evolutionary stages. This is supported by the H$\alpha$ emission and dust, seen in the HST images (see Figure\,\ref{fig:slit_alignment}). Source D with the highest ${E(B-V)_{\rm He\,II}}$ value is visibly affected by diffuse strong H$\alpha$ emission situated to its east. For sources A-C most of the dust and gas has already been displaced and a cavity has been created in the ISM by star formation feedback. However, their ${E(B-V)_{\rm Balmer}}$ and ${E(B-V)_{\rm UV}}$ values do not show such a strong variation. Overall these values are in good agreement with each other. This might be due to the fact that the spatial resolution of the MUSE observation is not enough to resolve each individual source (See Section~\ref{sssect:muse}).

\paragraph{Mrk\,33~}~This galaxy has two bright star clusters aligned within the slit but only source B has detected He\,II lines. The estimate of ${E(B-V)_{\rm Balmer}}$ was done with Balmer lines measured by SDSS with an aperture ($3^{\prime\prime}$) covering both sources. The UV and Balmer measurements suggest the absence of dust, whereas the He\,II measurement deviates from this with a dust attenuation of $E(B-V)_{\rm He\,II} = 0.17\pm0.06\,{\rm mag}$. By studying the morphology of Source B in Figure\,\ref{fig:slit_alignment}, we find that the cluster itself has no strong H$\alpha$ emission, indicating that star-formation feedback has already pushed away dust and gas. 

\paragraph{NGC\,3049~}~
We find good agreement between the He\,II and UV dust reddening estimates for Source A. Source B has no significant He\,II emission but we derived the $E(B-V)_{\rm UV}$ value of 0.42\,mag. This value is significantly higher then for Source A (0.13\,mag) and might be the reason for the non-detection of the He\,II lines due to dust attenuation. However, we do not see any traces of nebular H$\alpha$ emission around source B, which might be due to the fact that this source is already older than $\sim5\,{\rm Myr}$ and the W-R star population has already disappeared.  
The dust reddening estimated from SDSS Balmer-line measurements of $E(B-V)_{\rm Balmer}=0.29\,{\rm mag}$ is in between the UV estimated values of Source A and B. This might be the result of the blending of both sources within the SDSS fiber.

\paragraph{NGC\,3125~}~
Source A in NGC\,3125 is the most discrepant data point in Figure~\ref{fig:compare_ebv}. While the ${E(B-V)_{\rm Balmer}}$ determined from the Balmer lines and from He\,II agrees reasonably well, the ${E(B-V)_{\rm UV}}$ from the UV slope disagrees. The UV value of $E(B-V)_{\rm UV} = 0.53\,{\rm mag}$ significantly exceeds the results obtained via the other two methods. Inspection of thespectrum taken with the Cosmic Origin Spectrograph (COS) confirms a very red UV slope and therefore high reddening. \citet{chandar_ngc_2004} pointed out the unusually high equivalent width of He\,II\,$\lambda$1640 in NGC\,3125-A and proposed a scenario of a large population of W-R stars in conjunction with a dust morphology resulting in larger dust attenuation for the W-R population than for the OB population responsible for the stellar continuum. The peculiar shape of the UV continuum is evident in large-aperture ($20^{\prime\prime} \times 10^{\prime\prime}$) spectroscopic observations obtained with the International Ultraviolet Explorer \citep{kinney_atlas_1993}. The UV spectrum of NGC\,3125 is steeply rising from $2200\,{\rm \AA}$ to $1700\,{\rm \AA}$ by an essentially flat continuum at shorter wavelengths. The flat wavelength region is used for determining $\beta$ in the current work. This bimodal continuum shape is most pronounced in NGC\,3125 compared to other galaxies in the atlas of \citep{kinney_atlas_1993}

\citet{wofford_extreme_2023} found evidence of very massive stars having initial masses of up to 300 M$_{\odot}$ in NGC\,3125-A. Such stars mimic spectral features of classical W-R stars but are otherwise still core-hydrogen burning main-sequence stars. The detection of highly excited OV\,$\lambda$1371 by \citet{wofford_extreme_2023} is consistent with these stars being hot ($T{\rm _{eff} \approx 50, 000 K}$) and luminous ($L \approx 10^7 {\rm L_{\odot}}$). If so, they would be a significant source of hydrogen-ionizing photons. The additional supply of ionizing photons will enhance the nebular contribution to the spectral energy distribution. Of particular interest in this context is the wavelength region around 1500\,${\rm \AA}$, where the energy flux of the two-photon continuum peaks \citep{johnstone_hydrogen_2012}. A stronger two-photon continuum will redden the total (stellar + nebular) continuum. Consequently, the flat spectral slope observed in NGC\,3125-A could result from the presence of very massive stars whose ionizing photon output increases the nebular continuum in the UV.

%\paragraph{NGC\,4214~}~The broad He\,II emitter in NGC\,4214 is characterized by a well visible bubble of hydrogen which most likely formed from star formation feedback associated with this source as this is the most luminous source in this region (see Figure\,\ref{fig:slit_alignment}). 
\paragraph{NGC\,4214~}~The broad He\,II emitter in NGC\,4214 is characterized by a distinct bubble structure of hydrogen around this source. This is most likely formed from star formation feedback associated with this source as this is the most luminous source in this region (see Figure\,\ref{fig:slit_alignment}).
We find good agreement between all three measurements, keeping in mind that the measurement of the Balmer lines was taken for the entire galaxy \citep{moustakas_integrated_2006}. However, the H\,II region hosting the W-R stars is clearly the brightest source and might dominate the ionized gas spectra, resulting in similar $E(B-V)$ values.

\paragraph{NGC\,4670~}~Even though three sources were found in NGC\,4670, only Source A has significant He\,II emission. The SDSS spectroscopic fiber is centered on Source A, which is the brightest optical source in this region presumably dominating the SDSS spectrum. This source is also characterized by a negligibly small dust attenuation estimate whose values agree within the errors for the three methods. 

\paragraph{Tol\,1924-416~}~
This is a bright source with no significant direct neighbor. This makes it very suitable for our Balmer-line measurement with Archive MUSE observation. Interestingly this source appears to have ejected all its dusty envelope since we do not detect any dust attenuation with all three of the used methods.

\paragraph{Tol\,89~}~
This is the only source which has been observed with a Balmer line estimate of comparable aperture by \citet{sidoli_massive_2006}. 
The UV slope is very blue, indicative of little dust attenuation ($E(B-V){\rm _{UV} = 0.01\,mag}$), whereas the Balmer and He\,II line ratios indicate dust of 0.17 and 0.07\,mag, respectively. The morphology as seen in Figure\,\ref{fig:slit_alignment} does also not indicate a large dust content.

%Summarizing, regarding the three correlations of the {\it E(B-V)} values tested here, we find overall good agreement with each other. Even though NGC\,3125 shows extreme variation in its dust estimate, we find positive correlation coefficients for all three combinations of $E(B-V)_{\rm He\,II} \,vs.\, E(B-V)_{\rm Balmer} = 0.21$, $E(B-V)_{\rm UV} \,vs.\, E(B-V)_{\rm Balmer} = 0.47$ and $E(B-V)_{\rm UV} \,vs.\, E(B-V)_{\rm He\,II} = 0.48$. The less significant correlation between the He\,II and Balmer estimates might be due to the fact that not all Balmer estimates are probing the exact same regions as for He\,II and UV.

\subsection{Further application in different redshift regimes}
Two follow-up projects to our pilot analysis of broad He II lines are desirable: 
First, a more rigorous testing with higher statistics in the local Universe, accompanied by an extended study of the emitter's properties, would help to put this dust tracer for young ($<5\,{\rm Myr}$) stellar populations on a firmer footing. Subsequently, we can apply this method to targets at higher redshift to probe the dust reddening in starburst regions beyond the local Universe.  
In the following we will discuss available data and also the feasibility in different redshift regimes.

The challenge to observe broad He\,II lines in the local universe is the need for space-based spectroscopy for the He\,II\,$\lambda$1640\,\AA\ line. 
A well suited HST program is the COS Legacy Archive Spectroscopic SurveY (CLASSY), as it provides a UV spectral database of 45 nearby ($0.002 < z < 0.182$) galaxies with dust reddening values of $0.02 < E(B-V) < 0.67$ \citep{berg_cos_2022}. The COS on board of HST observes targets with a 2.5$^{\prime\prime}$ circular aperture. 
This survey has the distinct advantage that it provides high-quality (${\rm S/N_{\rm 1500\AA} \gtrsim 5/resel}$) and high-resolution (${\rm R\sim15,000}$) \citep{james_classy_2022} spectra which are ideal in order to search for broad He\,II emitters. 
CLASSY has a broad range of galaxy types which biased towards UV-bright star-forming galaxies. The survey covers stellar masses ranging over ${\rm log(M_*) \sim 6-10 M_{\odot}}$, star formation rates of ${\rm log(SFR) \sim-2}$ to ${\rm +2\,M_{\odot}\,yr^{-1}}$ and oxygen abundances of ${\rm 12+log(O/H) \sim 7-9}$ \citep{berg_cos_2022}.
The CLASSY sample provides also optical spectra for all targets covering the optical He\,II\,$\lambda$4686 line. The optical spectra are collected from archives including SDSS, VLT/VIMOS integral field unit (IFU), MMT Blue Channel Spectrograph, Keck/KCWI IFU, Keck/ESI and VLT/MUSE IFU. As discussed in \citet{arellano-cordova_classy_2022}, the SDSS and the COS have different apertures of 3$^{\prime\prime}$ and 2.5$^{\prime\prime}$, respectively. The spectra obtained with the MMT Blue Channel Spectrograph and KECK/ESI have a long-slit aperture with a width of 1$^{\prime\prime}$.
The spectra taken from IFUs on the other hand are extracted in a 2.5$^{\prime\prime}$ aperture.
Even though not all optical spectra were obtained with the exact same aperture as the one with COS, we can assume that broad He\,II lines are originating from only one dominating stellar population. Significant broad He\,II\,$\lambda$1640 has been reported in one galaxy (J0127--0619 or Mrk\,996)

% further UV samples from the local universe
The CLASSY sample stands out due to its good spectral coverage and sample size and is therefore well suited for an extended search for broad He\,II\,$\lambda$1640 emitters. \citet{senchyna_ultraviolet_2021} discuss a sample of ten local star-forming galaxies with available UV and optical spectroscopy. Broad W-R features have been detected in all ten galaxies. This sample could be an obvious extension of the sample of eight presented here.  
A systematic search for broad He\,II\,$\lambda$4686 lines in the SDSS DR\,6 has been performed in \citet{brinchmann_galaxies_2008} resulting in 570 galaxies with significant W-R line detection and further 1115 potential candidates. This sample can be used as a basis of future UV observations providing He\,II\,$\lambda$1640 line measurements.

For galaxies at a redshift $z\gtrsim1.0$ the He\,II\,$\lambda$1640 line is observable in the optical wavelength (for $z=1.0$ He\,II\,$\lambda$1640 is observed at $\sim 3300\,{\rm \AA}$) and can in principle be obtained from the ground. 
% experiment of high redshifted sources
In order to explore the detection of He\,II lines in galaxies at a higher redshift, we project observed fluxes  for the galaxy NGC\,3125 to higher distances. This galaxy is the brightest known He\,II\,$\lambda$1640 emitter in the local Universe \citep{chandar_ngc_2004}.  
This source has observed He\,II\,$\lambda$1640 and $\lambda$4686 line fluxes of $23.76\pm1.1 \times 10^{-15}{\rm ergs/s/cm^2}$ and $9.59\pm0.3 \times 10^{-15}{\rm ergs/s/cm^2}$, respectively. 
By using the relation between luminosity $L$ an observed flux $F$ 
\begin{equation}
    L = 4\pi\,F\,D_{\rm L},
\end{equation}
where $D_{\rm L}$ is the luminosity distance, we can assume that such a source with the same luminosity has a predicted flux at a given redshift of
\begin{equation}
    F(z) = \frac{F(z=0.0037) \times D_{\rm L}(z=0.0037)^2}{D_{\rm L}(z)^2}.
\end{equation}
This leads to predicted fluxes of $13.7 \times 10^{-20}{\rm ergs/s/cm^2}$ and $5.5 \times 10^{-20}{\rm ergs/s/cm^2}$ for the He\,II\,$\lambda$1640 and $\lambda$4686 lines, respectively, at a redshift of $z=1.0$. 
State-of-the-art spectrographs mounted on large telescopes like the VLT are, however, not capable of detecting the He\,II lines. 
As an example, we estimated the S/N values of the He\,II\,$\lambda$1640 and the $\lambda$4686 of $\sim10^-5$ per pixel for 1h VLT/XSHOOTER observation\footnote{We used the ESO exposure time calculator \url{https://www.eso.org/observing/etc/}}, as this instrument provides the needed spectral coverage and resolution. 

By stacking spectra broad He\,II\,$\lambda$1640 W-R features were observed in $z\sim3$ Lyman break galaxies \citep[$N\sim 1000$;][]{shapley_rest-frame_2003}, in star-forming galaxies at $z = 2.40\pm0.11$ \citep[$N\sim 30$; ][]{steidel_reconciling_2016}, and in highly magnified gravitationally lensed galaxies at redshifts
$1.6 < z < 3.6$ \citep[$N\sim 14$; ][]{rigby_magellan_2018}. 
In \citet{saxena_properties_2020}, 6 broad He\,II\,$\lambda$1640 emitters at a mean redshift of $z=2.7$ using VLT/VIMOS were detected with fluxes between 2.0 and 23.1 $\times 10^{-18}{\rm ergs/s/cm^2}$. 
However, given the fact that these observations only have a S/N ratio between 2 and 7 and that for these sources the He\,II\,$\lambda$4686 is shifted into the near IR wavelengths, it is likely that this line remains undetectable at these redshifts from the ground. With the current technology it is not possible to make a positive prediction at what distance the He\,II line ratios can be used as dust-reddening tracers. Therefore, the most promising strategy would be to first extend the limits in the local Universe and test this method using  targets up $\sim$ 100~Mpc using space based UV observations.

\section{Conclusion}\label{sect:conclusion}
We discuss a novel method to determine the dust attenuation in star-forming galaxies containing W-R stars. We applied the stellar He\,II\,$\lambda$1640 to $\lambda$4686 line ratio as a dust-reddening tracer to a sample of eight galaxies. We measured the flux of both lines with HST STIS long-slit observations in the UV and optical, respectively. Both measurements were taken for the same position and aperture to guarantee co-spatial observations. 
As this is the first application of this method, we compared our dust-reddening estimates to two commonly used methods. We measured the UV slope from HST STIS observations and collected Balmer-line measurements for all eight galaxies from the literature.
We find consistency between the three methods except for NGC\,3125, which we interpret as due to an exceptional contribution from a nebular continuum affecting the UV continuum. Even though the Balmer line measurements are not always co-spatial with the HST STIS observations, we still find a significant correlation with the UV and He\,II methods. This is most likely due to the fact that the selected sources for this study are all bright young star clusters which dominate the luminosity of their local environment. We find tentative evidence for off-sets between the reddening values derived from the three methods: the He II method suggests the largest values, with the UV method leading to the lowest reddening. As these dust-reddening estimates probe different age regimes, we may witness different stages of dust clearing. This pilot study is still limited by small-number statistics, and an expanded data set is needed. Suitable data sets exist in the local universe, and there are opportunities for extending this method to more galaxies at larger distance.

% references for software
\software{astropy \citep{astropy_collaboration_astropy_2022, astropy_collaboration_astropy_2018, astropy_collaboration_astropy_2013},
          specutils \citep{earl_astropyspecutils_2023}
          calstis \citep{sohn_stis_2019},
          matplotlib \citep{hunter_matplotlib_2007},
          numpy \citep{harris_array_2020}
          }

\section*{Acknowledgements}
This research is based on observations made with the NASA/ESA Hubble Space Telescope obtained from the Space Telescope Science Institute, which is operated by the Association of Universities for Research in Astronomy, Inc., under NASA contract NAS 5–26555. These observations are associated with programs  9036, 7513, 15846, 06580, 11146, 11987, 10400, 11360, 06639 and 06708. 
Support for this work has been provided by NASA through grant No. AR-15036.

Based on observations collected at the European Southern Observatory under ESO programs 095.B-0321 (PI: Vanzi), 094.B-0745 (Garcia-Benito, Ruben) and 60.A-9314 (MUSE Science Verification).

This research has made extensive use of the NASA/IPAC Extragalactic Database (NED) which is operated by the Jet Propulsion Laboratory, California Institute of Technology, under contract with the National Aeronautics and Space Administration.

Funding for the Sloan Digital Sky 
Survey IV has been provided by the 
Alfred P. Sloan Foundation, the U.S. 
Department of Energy Office of 
Science, and the Participating 
Institutions. 

SDSS-IV acknowledges support and 
resources from the Center for High 
Performance Computing  at the 
University of Utah. The SDSS 
website is www.sdss4.org.

SDSS-IV is managed by the 
Astrophysical Research Consortium 
for the Participating Institutions 
of the SDSS Collaboration including 
the Brazilian Participation Group, 
the Carnegie Institution for Science, 
Carnegie Mellon University, Center for 
Astrophysics | Harvard \& 
Smithsonian, the Chilean Participation 
Group, the French Participation Group, 
Instituto de Astrof\'isica de 
Canarias, The Johns Hopkins 
University, Kavli Institute for the 
Physics and Mathematics of the 
Universe (IPMU) / University of 
Tokyo, the Korean Participation Group, 
Lawrence Berkeley National Laboratory, 
Leibniz Institut f\"ur Astrophysik 
Potsdam (AIP),  Max-Planck-Institut 
f\"ur Astronomie (MPIA Heidelberg), 
Max-Planck-Institut f\"ur 
Astrophysik (MPA Garching), 
Max-Planck-Institut f\"ur 
Extraterrestrische Physik (MPE), 
National Astronomical Observatories of 
China, New Mexico State University, 
New York University, University of 
Notre Dame, Observat\'ario 
Nacional / MCTI, The Ohio State 
University, Pennsylvania State 
University, Shanghai 
Astronomical Observatory, United 
Kingdom Participation Group, 
Universidad Nacional Aut\'onoma 
de M\'exico, University of Arizona, 
University of Colorado Boulder, 
University of Oxford, University of 
Portsmouth, University of Utah, 
University of Virginia, University 
of Washington, University of 
Wisconsin, Vanderbilt University, 
and Yale University.

\bibliographystyle{aasjournal}
\bibliography{bibliography}

\end{document}